\documentclass{mem}
\usepackage{natbib}\usepackage{txfonts}\usepackage{balance}
\usepackage{graphicx}
\usepackage[a4paper,breaklinks,dvipdfm]{hyperref}
\idline{75}{282}
\begin{document}
\def\teff{$T\rm_{eff }$}
\def\kms{$\mathrm {km s}^{-1}$}

\title{
The Search for SW Sex Type Stars
}
   \subtitle{}
\author{
Linda Schmidtobreick\inst{1} 
\and
Pablo Rodr\'\i guez-Gil\inst{2,3,4}
\and Boris T. G\"ansicke\inst{5}
          }
  \offprints{L. Schmidtobreick}
\institute{
European Southern Observatory,
Alonso de Cordova 3107,
Vitacura, Santiago, Chile
\and
Instituto de Astrof\'\i sica de Canarias, 
V\'\i a L\'actea,
Santa Cruz de Tenrife, Spain 
\and 
Departmento de Astrof\'\i sica, 
Universidad de La Laguna, 
Santa Cruz de Tenerife, Spain
\and
Isaac Newton Group of Telescopes, 
Apartado de correos 321, 
38700 Santa Cruz de la Palma, Spain
\and
Department of Physics, 
University of Warwick, 
Coventry CV4 7AL, UK
}

\authorrunning{Schmidtobreick et al.}

\titlerunning{The Search for SW Sex Type Stars}

\abstract{
All eclipsing nova-likes in the 2.8--4\,h orbital period 
range belong to the group of SW\,Sex stars, and as such experience
very high mass transfer rates. Since the physical properties of a star
should be independent of the inclination it is observed at, 
this suggests that all
or at least a large fraction of the
non- or weakly-magnetic cataclysmic variables in this period range
are physically SW\,Sex stars. 

We here present preliminary results of a large campaign to search
for SW\,Sex characteristic features in the spectra of such stars.
We find that 14 out of the 18 observed non-eclipsing 
cataclysmic variables belong to the group of SW\,Sex stars the
classification of the other four is uncertain from our data. This confirms the
domination of SW\,Sex stars in the period range of 2.8--4\,h
just above the period gap. Since all long-period systems
need to cross this range before entering the gap, the SW\,Sex phenomenon
is likely to be an evolutionary stage in the life of a cataclysmic variable.
}
\maketitle{}

\section{Introduction}
SW\,Sextantis stars are a sub-group of cataclysmic variables (CVs) that
were originally characterised by \cite{thorstensenetal91-1}.
They were defined as eclipsing nova--like stars with
high velocity, emission line wings extending up to 4000\,km/s, 
inconsistent with an origin in a standard accretion disc. They show narrow
absorption features in the Balmer and He\,{\sc i} lines near the inferior
conjunction of the white dwarf, and large orbital phase offsets
($\sim 0.2$ cycle) of
the radial velocity curves with respect to the photometric ephemeris
\citep{rodriguez-giletal07-1}.
They also have large absolute magnitudes and hot white dwarfs, 
implying extremely high accretion rates exceeding the expected 
rates based on standard magnetic braking as angular momentum
loss mechanism \citep{Town+09}.

While SW\,Sex stars were considered rare objects at the beginning, later
surveys (see \cite{2005ASPC..330....3G} for an overview) have shown 
that a surprisingly large number of the newly identified systems 
are deeply eclipsing SW\,Sextantis stars
with orbital periods between 2.8\,h and 4\,h. Populating the upper edge 
of the orbital period gap makes them very interesting objects
in the context of CV evolution. In detail, 13 out of 48
nova-like systems in this period range -- these are all the eclipsing ones --
belong to the sub-class of SW\,Sex stars.

Since being an eclipsing system is not an intrinsic
physical property of the star but rather depends on the angle under which
the binary is observed, it appears entirely plausible that all
non- or weakly-magnetic CVs just above the period gap 
are physically SW\,Sex stars, i.e. experience a very high mass transfer rate. 
This would be
of major significance for the evolutionary theory of CVs, as all long-period
CVs have to pass through this range before 
entering the period gap. 

During the last years, we have conducted a project to test the 
hypothesis that all
nova--like stars in the 2.8--4\,h period regime are physical SW\,Sex type
stars, even if they are not eclipsing. Examples for such stars are e.g.
RR\,Pic \citep{schmidto+03} or V533\,Her \citep{2002MNRAS.337..209R}.
We selected a sample of candidates: non-eclipsing nova-likes and 
old novae with an orbital period of 2.8--4\,h that are sufficiently bright
to perform time-series spectroscopy. We aim to analyse the emission 
lines of these 
stars searching for the presence of SW\,Sex characteristics such as
broad line wings with large--amplitude radial velocity
variations, single--peaked line profiles with phase-dependent central
absorption, and phase lags between the radial velocity modulation in the line
cores and wings. We also check for line flaring, an additional feature 
that is often observed in
SW\,Sex stars but also in intermediate polars and that is manifested in  
fast oscillations of the emission line flux and velocity with periods 
around 10--20\,min.
First results of this campaign are discussed in 
\cite{2007MNRAS.374.1359R}.

\section{Data}
In total, we have observed 18 non-eclipsing CVs with periods between 
2.8\,h and 4\,h. Time series spectroscopy
was done covering at least one orbit but most of the stars were followed
over longer time ranges covering up to two orbits. The data of the first
set are described in \cite{2007MNRAS.374.1359R}.

The new set of candidate SW\,Sex stars were observed in 
January and December 2009 with EFOSC2 \citep{buzz+84}
mounted at the
NTT of the European Southern Observatory on La Silla, Chile.
Grism \#\,20 centred on H$\alpha$ was used to perform time-series
spectroscopy covering at least one orbit for each system. The data 
were reduced using standard procedures in {\tt PAMELA\footnote{Tom Marsh's
packages PAMELA and MOLLY  are available at http://deneb.astro.warwick.ac.uk/ phsaap/software/}} and {\tt MOLLY$^1$}. The 
wavelength calibration yielded a final FWHM resolution of 3.8\,\AA. All
further analysis was done using {\tt MIDAS\footnote{MIDAS is distributed by ESO at http://www. eso.org/sci/software/esomidas/}}.
\section{Preliminary Results}
\begin{figure}
\resizebox{\hsize}{!}{\includegraphics[clip=true]{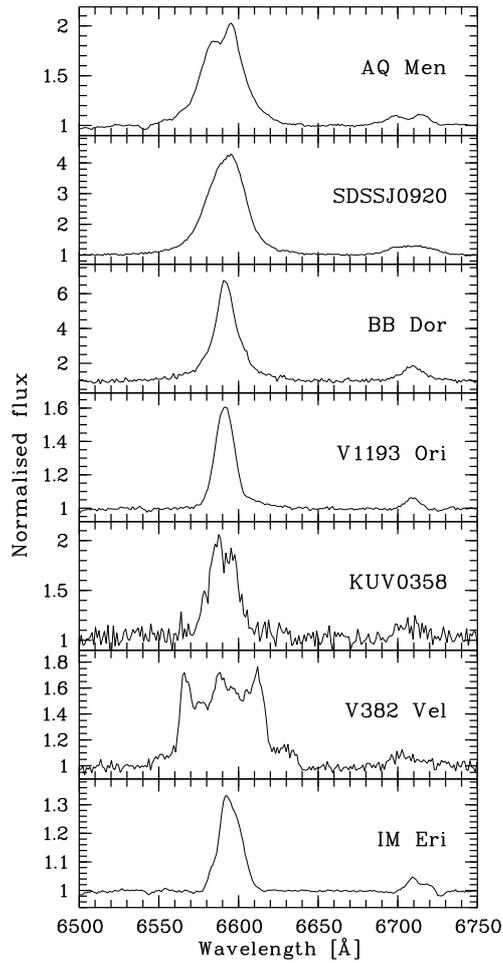}}
\caption{\footnotesize Averaged normalised spectra of the new set of
SW\,Sex candidate stars.}
\label{plot_09}
\end{figure}

In Figure \ref{plot_09} the average spectra of the new candidates are
plotted. All of them are dominated by H$\alpha$ in emission and also show
the He\,I\,$\lambda$6678 line. As expected for SW\,Sex stars, most 
of them show single peak emission instead of the characteristic 
double-peak profile of lines originating in an accretion disc.
The exceptions are AQ Men which shows a weak double-peak profile, 
KUV0358 which seems to show several components although part of them might
rather be noise, and V382\,Vel which still shows the shell of its 
1999 nova outburst via
two widely separated emission peaks bracketing the single-peaked 
emission line from the central binary.
\begin{figure}
\resizebox{!}{5.5cm}{\includegraphics[clip=true]{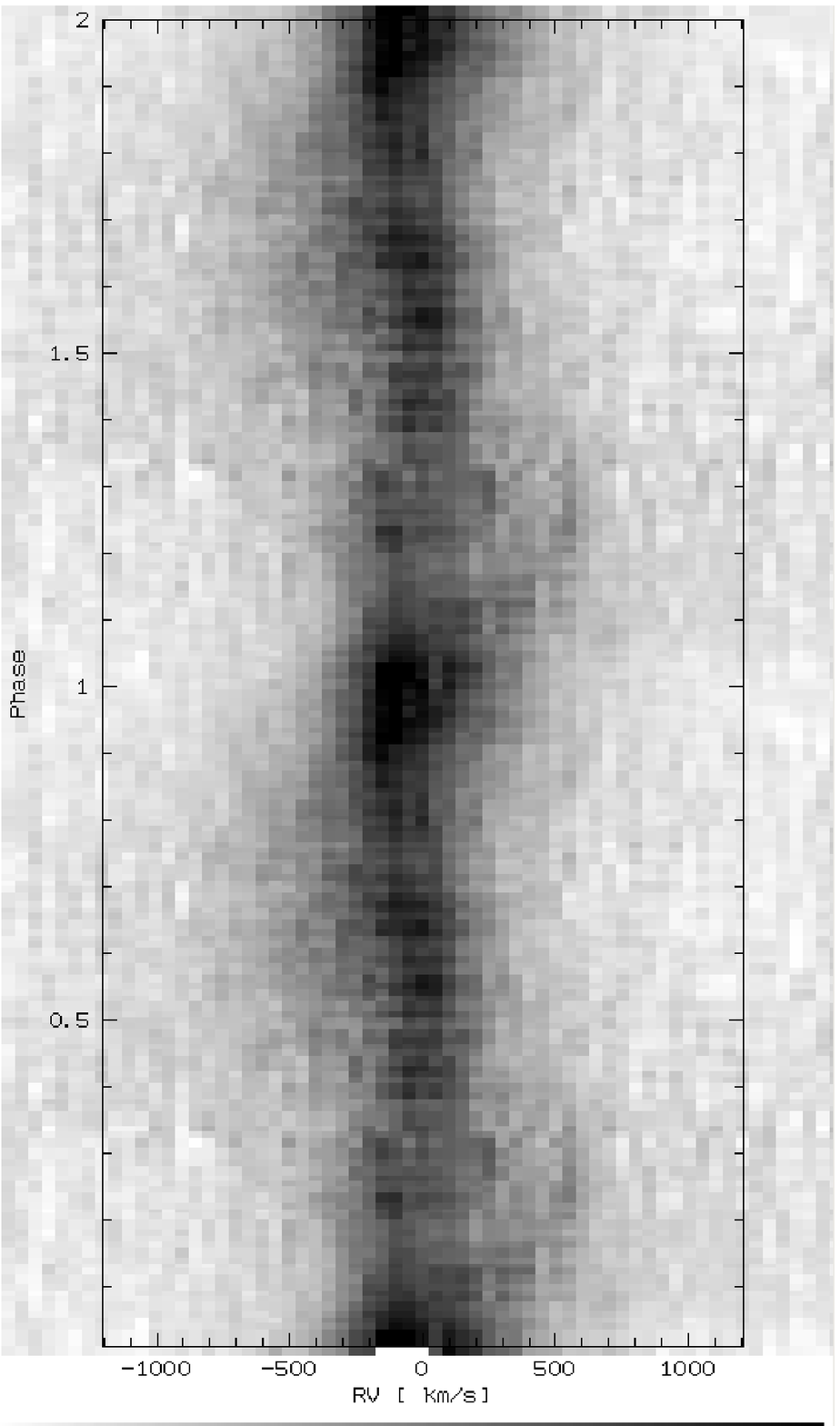}}\, 
\resizebox{!}{5.5cm}{\includegraphics[clip=true]{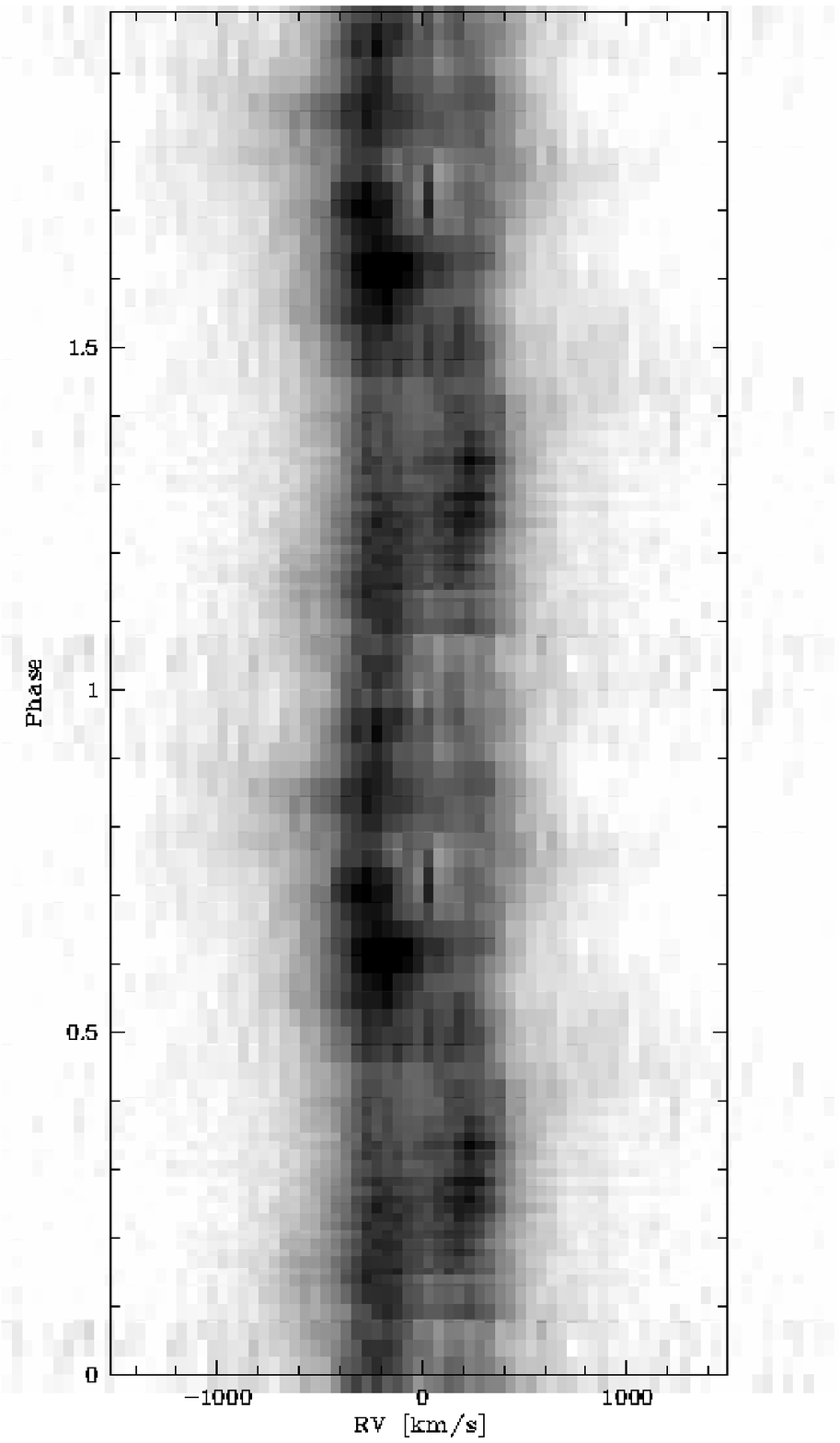}}
\caption{\footnotesize Trailed spectra of the H$\alpha$ line of BB\,Dor 
(left) and KUV03580+0614 (right). The phase for BB\,Dor is determined using
the ephemeris of \cite{rod+2011} which has sufficient accuracy to be
extrapolated to our observations. The orbital period $P=0.143$\,d 
for KUV03580+0614 was taken from \cite{2001PASP..113.1215S} the zero
phase was set arbitrary.
}
\label{trail}
\end{figure}

A first analysis of the radial velocities and trailed spectra shows
that all new systems seem to experience the high velocity S--wave, one 
of the main characteristics of SW\,Sex stars 
(see Fig.\,\ref{trail} for examples). Also a central absorption 
feature at a specific phase is present for all these stars.
For V382\,Vel, these results are
less secure as the lines from the nova shell muddle the outer wings of the
central line and a careful analysis is needed for confirmation.
 The phase lag has not yet been found
for all stars mainly because a more thorough study of the
different components in the emission line is needed to determine
the correct zero-phase. Also the flaring has not yet been checked for in
any of the new stars. We leave these results for a later paper.
Still, with the presence of the S--wave as well as the central absorption
we are pretty confident that all our new candidates are in fact 
SW\,Sex stars.

Thus, from the 18 observed nova-likes in the orbital period range of 2.8--4\,h,
14 belong to the group of SW\,Sex stars, see Table \ref{candidates} for 
details. The data for V849\,Her and V393\,Hya are rather poor and do not 
allow a conclusive classification. LQ\,Peg is of very low inclination and
much higher spectral resolution is needed to find any features in the 
emission line. The spectrum of V992\,Sco is still
dominated by the nova outburst and does not yet show any signature of the
binary.

\begin{table*}
\caption{The observed SW\,Sex characteristics and the final classification for
all candidates}
\label{candidates}
\begin{center}
\begin{tabular}{|l| c| c| c| c| c| c|}
\hline
System& single peak&S--wave&0.5--abs.&phases& flaring &SW\,Sex? \\
\hline
\hline
HL\,Aqr$^*$ & $\surd ^a$& $\surd$ & $\surd$ & $\surd$ &  x & yes\\
BO\,Cet$^*$ & $\surd$ &  $\surd$ & $\surd$ & $\surd$ &  $\surd$ & yes \\
BB\,Dor$^{**}$ & $\surd$ & $\surd$ & $\surd$ & $\surd$ & ? & yes \\
IM\, Eri$^{**}$ & $\surd$ & $\surd$ & $\surd$ & $\surd$ & ? & yes \\
V849\,Her$^*$& $\surd$ & x & x & x &      x&      no $^c$\\
V393\,Hya$^*$&      x&      x&      x&      x&      x&     no $^c$\\
AQ\,Men$^{**}$ & x & $\surd$ & $\surd$ & ? & ? & yes \\
AH\,Men$^*$ & $\surd$ & $\surd$ & $\surd$ &  $\surd$ &  x & yes \\
KQ\,Mon$^{**}$ & $\surd$ & $\surd$ & $\surd$ &  $\surd$ &  $\surd$ & yes \\
V380\,Oph$^*$& $\surd$ & $\surd$ & $\surd$ & $\surd$ &  $\surd$ & yes \\
V1193\,Ori$^{**}$& $\surd$ & $\surd$ & $\surd$ &  $\surd$ &  ? & yes \\
LQ\,Peg$^*$ & $\surd$  &   x   &    x&      x&      x&      no $^c$\\
AH\,Pic$^*$ &  $\surd ^a$  & $\surd$ &x &$\surd$ &x &yes \\
V992\,Sco$^*$&      -&      -&      -&      -&       -&      - $^b$\\
LN\,UMa$^*$ & $\surd$ &$\surd$ &$\surd$ & $\surd$ & x & yes  \\
V382\,Vel$^{**}$ & x$^b$ & $\surd$ & $\surd$ & $\surd$ & ? & yes \\
SDSS\,J0920+0043$^{**}$ & $\surd$ & $\surd$ & $\surd$ & $\surd$ & ? & yes \\
KUV\,03580+0614$^{**}$ & x & $\surd$ & $\surd$ & $\surd$ & ? & yes \\
\hline
\end{tabular}
\\
\parbox{11cm}{\footnotesize
$^a${in absorption.};
$^b${spectrum still dominated by shell emission from nova explosion};\\
$^c$ need higher quality data\\
$^*$ \cite{2007MNRAS.374.1359R};
$^{**}$\cite{schmidto+11}
}
\end{center}
\end{table*}

\section{Conclusions}
With the results so far, we have shown that the mayority of the 
nova-like stars in the 2.8--4\,h period range are of  
SW\,Sex nature. There might be the one or other oddball among the CVs
in this range, but the population in clearly dominated by SW Sex stars.
On the other hand, due to angular momentum loss, a CV that started with 
a long orbital period
%that during their evolution, the CVs that started with a long orbital period 
will eventually cross the 2.8--4\,h period range.
%before entering the period gap.
Our findings imply that these CVs will turn into an SW\,Sex star during this
phase of their lifetime. Therefore, the SW\,Sex phenomenon can be no longer 
considered a feature of individual stars but rather an evolutionary stage
of the CV population.
\begin{acknowledgements}
We acknowledge the use of Tom Marsh's {\tt PAMELA} and {\tt MOLLY} 
packages. We also made thorough use of the {\tt SIMBAD} database operated at 
CDS, Starssbourg, France. The data were collected at ESO, La Silla, Chile 
under program IDs 082.D-0138 and 084.D-0646. 
\end{acknowledgements}

\bibliographystyle{aa}

\end{document}